\documentclass[report]{amsart}
\usepackage{framed}         
\usepackage{soul}           
\usepackage{lmodern} 

\usepackage{rotating}
\usepackage{floatpag}
\rotfloatpagestyle{empty}
\usepackage{amsmath}
\usepackage{amsthm}
\usepackage{graphicx}

\usepackage{cite}
\usepackage{pstool}
\usepackage{chapterbib}

\usepackage{mathrsfs} 
\usepackage{amsfonts}
\usepackage{subcaption}
\usepackage{longtable}

\usepackage{mathtools,bm}  

\begin{document}

\title{A chart for the energy levels of the square quantum well}
\author{M. Chiani} 
\address{Dept. DEI ``G. Marconi'' \\ University of Bologna, Italy} 
\email{marco.chiani@unibo.it}

\date{September 29, 2016}

\begin{abstract}
A chart for the quantum mechanics of a particle of mass $m$ in a one-dimensional potential well of width $w$ and depth $V_0$ is derived. The chart is obtained by normalizing energy and potential through multiplication by ${8 m}{w^2} / h^2$, and gives directly the allowed couples (potential, energy), providing insights on the relation between the parameters and the number of allowed energy levels. 
\end{abstract}
\maketitle

\section{Introduction}

One of the classical examples 
 in quantum mechanics is the study of the allowed energy $E$ for a particle of mass $m$ under the effect of a potential well of width $w$ and finite potential depth $V_0$ (Fig.~\ref{fig:well}).
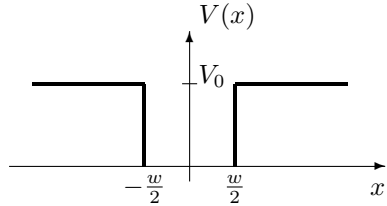
\begin{figure}[h]
\centering
\unitlength=1.00mm
\begin{picture}(59,30)
\put(0,5){\vector(1,0){50}}
\put(24,3){\vector(0,1){20}}
\put(29,25){\makebox(0,0)[cc]{$V(x)$}}
\put(27,17){\makebox(0,0)[cc]{$V_0$}}
\put(49,2){\makebox(0,0)[cc]{$x$}}
\put(30,2){\makebox(0,0)[cc]{$\frac{w}{2}$}}
\put(18,2){\makebox(0,0)[cc]{$-\frac{w}{2}$}}
\put(23,16){\line(1,0){2}}
\linethickness{0.4mm}
\put(18,5){\line(0,1){11}}
\put(30,16){\line(0,-1){11}}
\put(30,16){\line(1,0){15}}
\put(18,16){\line(-1,0){15}}
\end{picture}
%
\caption{Potential in a quantum well}
\end{figure}
\label{fig:well}
%
By solving the Schr\"{o}dinger equation it results that there is no quantization for $E>V_0$, while 
 for $E \leq V_0$ the quantized energy levels 
 are the roots of some transcendental equations. The graphical discussion of the allowed energy levels is commonly based on plots which must be drawn specifically for the numerical values of the potential depth $V_0$ of interest \cite{LanLif:65,Pit:55,Can:71,Spr:92,AroStr:00,Gri:B05,BonGri:06,Bin:15,Naq:15,Bar:15}.
 
In this note we show that, with a suitable normalization, the inverse problem $V_0=V_0(E)$ is in simple closed form, and can be represented with 
 a graph valid for arbitrary $V_0, m$ and $w$. The graph gives the allowed couples (potential, energy), with direct insights on energy quantization and on the number of energy levels. 

\section{Energy levels for the potential well}
The material in this section is well known and reported here just for the sake of completeness. 

The bounded allowed energy levels $E\leq V_0$ for a particle of mass $m$ under the effect of a potential well of finite width $w$ and potential $V(x)$ are given by the solutions of the time-independent Schr\"{o}dinger equation \cite{Gri:B05,Bin:15}
\begin{equation}\label{eq:sch}
-\frac{\hbar^2}{2m}  \frac{\partial^2 \psi(x)}{\partial x^2} + V(x) \psi(x) = E \, \psi(x) 
\end{equation}
where $\psi(x)$ is the particle wave functions, and $\int_{-\infty}^{\infty} |\psi(x)|^2 dx=1$. 
%
%
%
By introducing the normalized potential depth and energy
\begin{equation}
\tilde{V}_0= V_0  \frac{8 m}{h^2} w^2 \qquad \tilde{E}= E  \frac{8 m}{h^2} w^2
\end{equation}
%
%
%
where $h$ is the Planck constant, equation \eqref{eq:sch} for the quantum well is written as
\begin{equation}\label{eq:TISEnorm}
\frac{\partial^2 \psi(x)}{\partial x^2}=\left\{\begin{array}{lr}
  \displaystyle\left(\tilde{V}_0-\tilde{E}\right) \frac{\pi^2}{w^2} \psi(x)  &  |x| > w/2 \\
 \displaystyle -\tilde{E} \frac{\pi^2}{w^2}  \, \psi(x) &  |x| < w/2 \,.
\end{array}
\right.
\end{equation}
Due to symmetry of the potential, only even parity solutions $\psi(-x)=\psi(x)$ or odd parity solutions $\psi(-x)=-\psi(x)$ are possible. 



%
The even parity general solution of \eqref{eq:TISEnorm} is 
%
\begin{equation}\label{eq:psieven}
 \psi(x)=\left\{\begin{array}{lr}
  \displaystyle a \, \exp\left(- \pi\sqrt{\tilde{V}_0-\tilde{E}}  \, \frac{|x|}{w} \right)  &  |x| > w/2 \\
 \displaystyle b \cos\left(\pi \sqrt{\tilde{E}}  \, \frac{x}{w} \right) &  |x| < w/2 \,
\end{array}
\right.
\end{equation}
where $a, b$ are constants. To have continuity of $\psi(x)$ and its derivative at $\pm w/2$, the following equations have to be satisfied: 
\begin{align}
a &= b \cos\left( \frac{\pi}{2} \sqrt{ \tilde{E}}  \right) \exp\left(\frac{\pi}{2}\sqrt{\tilde{V}_0-\tilde{E}} \right)  \label{eq:tra1}\\
\sqrt{\tilde{V}_0-\tilde{E}} &= \sqrt{\tilde{E}}\tan\left(\frac{\pi}{2} \sqrt{\tilde{E}} \right)  \,. \label{eq:tra2}
 \end{align}
%

%
%
Similarly, the odd parity general solution of \eqref{eq:TISEnorm} is 
\begin{align}
\psi(x)=\left\{\begin{array}{lr}
 \displaystyle a \exp\left(- \pi \sqrt{\tilde{V}_0-\tilde{E}}  \, \frac{|x|}{w} \right) \text{sign}(x)  & |x| > w/2 \\
 \displaystyle b \sin\left(\pi \sqrt{\tilde{E}}  \, \frac{x}{w} \right)  & |x| < w/2 \,.
\end{array}
\right.
 \end{align}
Imposing the continuity of $\psi(x)$ and of its derivative at $\pm w/2$ we have
\begin{align}
a &= b \sin\left( \frac{\pi}{2} \sqrt{ \tilde{E}}  \right) \exp\left(\frac{\pi}{2}\sqrt{\tilde{V}_0-\tilde{E}} \right) \\
\sqrt{\tilde{V}_0-\tilde{E}} &= - \sqrt{\tilde{E}}\cot\left(\frac{\pi}{2} \sqrt{\tilde{E}} \right) \,.  \label{eq:tra4}
 \end{align}
In particular, equations \eqref{eq:tra2} and \eqref{eq:tra4} give the allowed energy levels for a fixed potential depth, for even and odd parity wave functions, respectively. 

\begin{figure}[h]
\centering
\includegraphics[width=0.5\columnwidth,draft=false,clip]{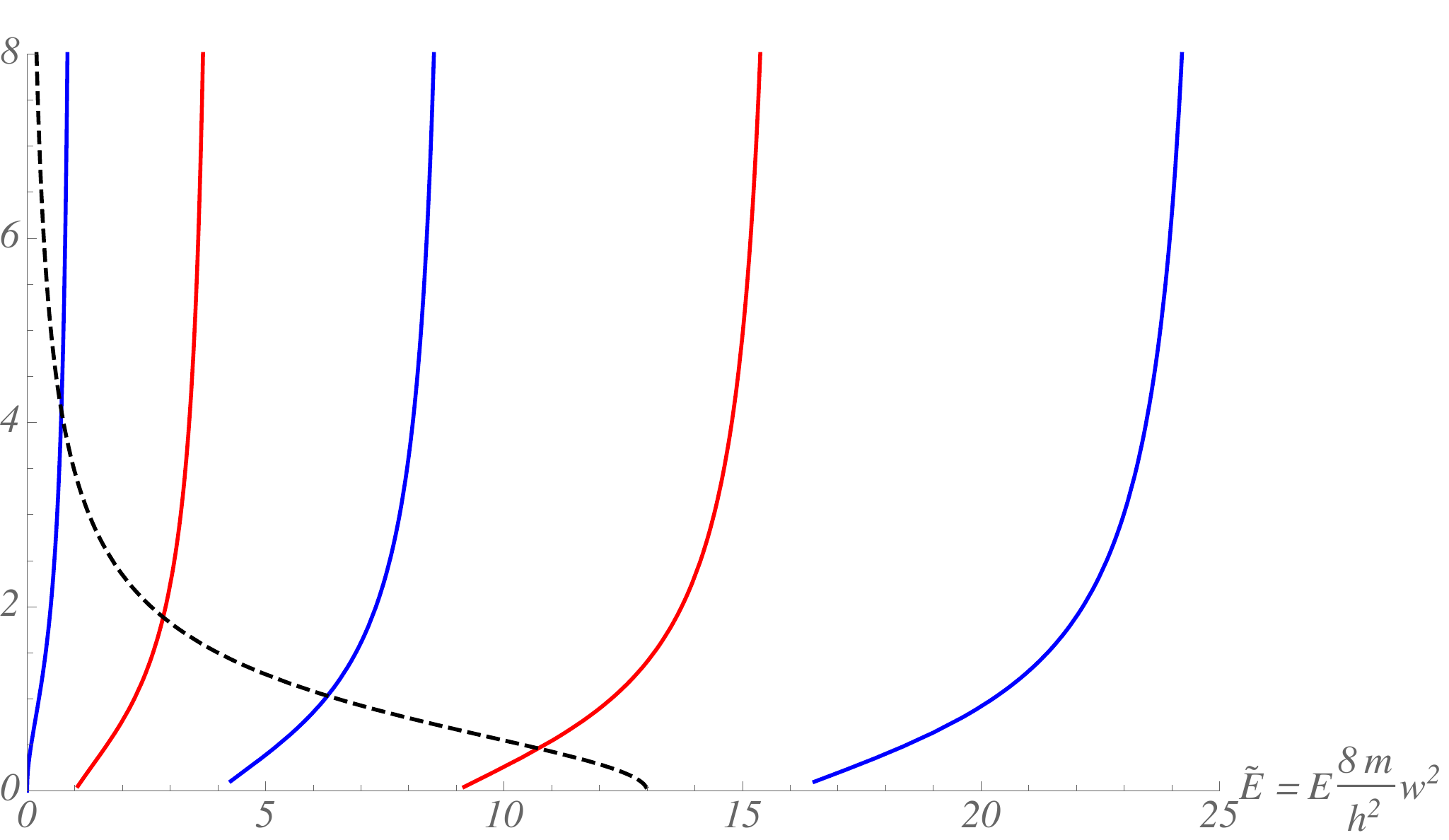}
\caption{Blue: the function $\tan(\sqrt{\tilde{E}} \, {\pi}/{2})$. Red: the function $-\cot(\sqrt{\tilde{E}} \, {\pi}/{2})$. Black: the function $\sqrt{\tilde{V}_0/\tilde{E}-1}$ for $\tilde{V}_0=13$. The allowed energy levels can be read as the abscissa of the intersections.}
\label{fig:classic}
\end{figure}

The transcendental equations \eqref{eq:tra2} and \eqref{eq:tra4} are usually discussed plotting separately the functions $\tan(\sqrt{\tilde{E}} \, {\pi}/{2})$, $-\cot(\sqrt{\tilde{E}} \, {\pi}/{2})$ and $\sqrt{\tilde{V}_0/\tilde{E}-1}$ for a given $\tilde{V}_0$ \cite{Gri:B05,Bin:15}. An example is reported in Figure~\ref{fig:classic}, for $\tilde{V}_0=13$. The abscissa of the intersections are the allowed energy levels. Unfortunately the graph requires drawing the curve $\sqrt{\tilde{V}_0/\tilde{E}-1}$ for the specific potential $\tilde{V}_0$.

\newpage
\section{A universal graph giving the couples (potential, energy)}
In order to provide a universal graph, valid for all cases, we look explicitly for the function $\tilde{V}_0\left(\tilde{E}\right)$. 
To this aim, we square \eqref{eq:tra2}, getting\footnote{Here the normalization makes the results easier to interpret with respect to \cite{Gue:72}.}
\begin{align}\label{eq:v0even1}
\tilde{V}_0-\tilde{E}= \tilde{E}\tan^2\left( \frac{\pi}{2}\sqrt{\tilde{E}}  \right)   \,.
 \end{align}
Thus, the inverse relation $\tilde{V}_0=\tilde{V}_0\left(\tilde{E}\right)$ for the even parity solutions is simply 
\begin{align}\label{eq:v0even}
{\tilde{V}_0= \tilde{E}\left(1+\tan^2\left( \frac{\pi}{2}\sqrt{\tilde{E}}  \right) \right)=\frac{{\tilde{E}}}{\cos^2\left(\sqrt{\tilde{E}} \, {\pi}/{2}\right)} = 
{{\tilde{E}}}{\, \sec^2\left( \frac{\pi}{2}\sqrt{\tilde{E}} \right)}} 
 \end{align}
where only the solutions with positive derivative have to be considered.\footnote{These are the only allowed for obvious physical reasons. In fact, it can be checked that the solutions of \eqref{eq:v0even} with negative derivative do not satisfy \eqref{eq:tra2}.} 

Similarly, for the odd parity case, from \eqref{eq:tra4} we get
\begin{align}
\tilde{V}_0-\tilde{E}= \tilde{E}\cot^2\left( \frac{\pi}{2}\sqrt{\tilde{E}}  \right)   \,.
 \end{align}
Thus, for the odd-parity wave functions the allowed energy levels are the solutions with positive derivative of the equation
\begin{align}\label{eq:v0odd}
{\tilde{V}_0= \tilde{E}\left(1+\cot^2\left( \frac{\pi}{2}\sqrt{\tilde{E}}  \right) \right)=\frac{{\tilde{E}}}{\sin^2\left( \sqrt{\tilde{E}} \, {\pi}/{2} \right)} =
{{\tilde{E}}}{\, \csc^2\left( \frac{\pi}{2}\sqrt{\tilde{E}} \right)} \,.  }
 \end{align}
%

\begin{figure}[h]
\centering
\includegraphics[width=0.99\columnwidth,draft=false,clip]{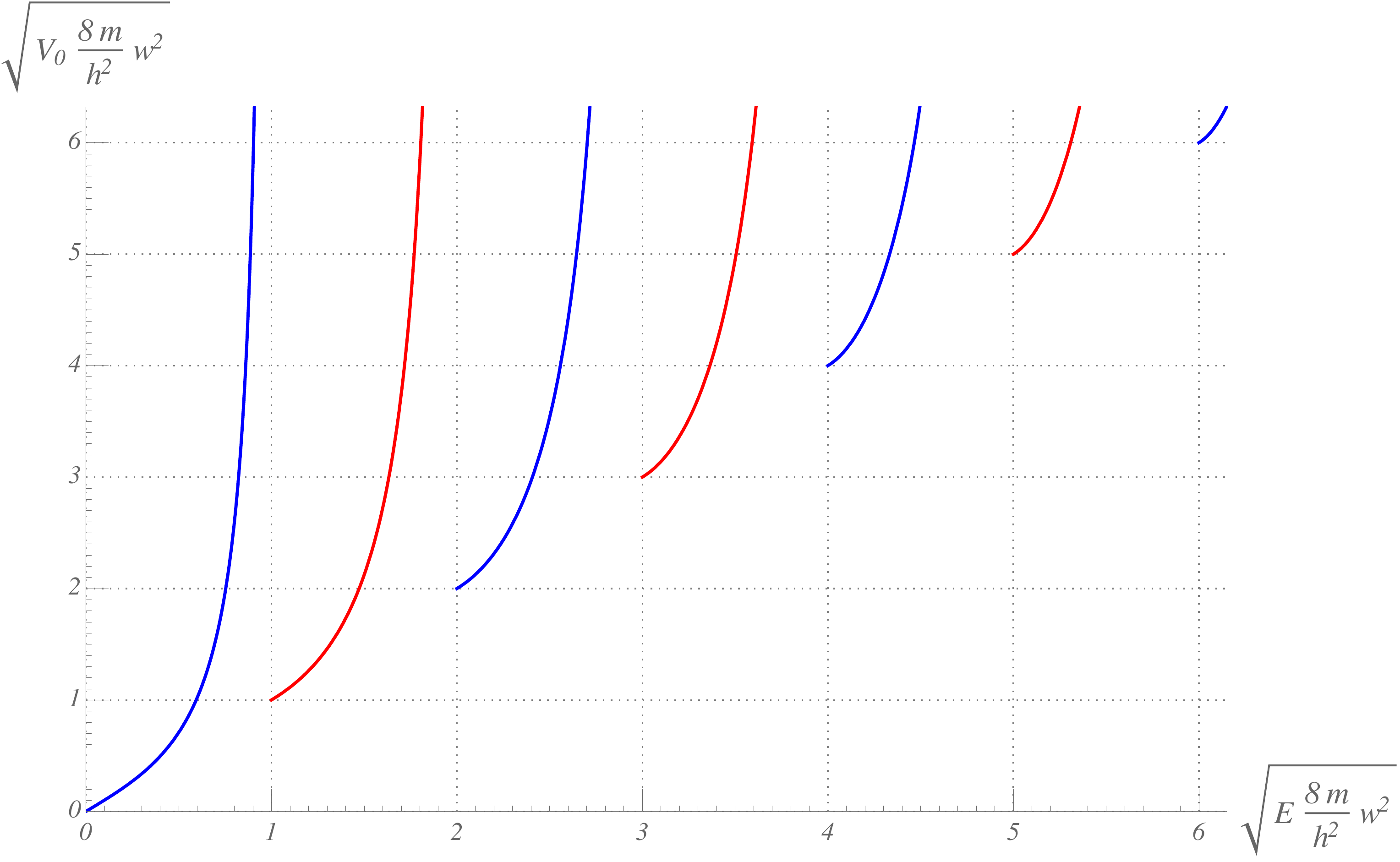}
\caption{Normalized potential depth vs. energy levels for a particle of mass $m$ and energy $E$ in a finite potential well of depth $V_0$ and width $w$. \\ 
Blue: plot of the function ${\sqrt{\tilde{E}}}\, |{\sec(\sqrt{\tilde{E}} \, {\pi}/{2})}|$ (positive derivative parts only) corresponding to the even parity solutions; \\
 Red: plot of the function ${\sqrt{\tilde{E}}}\, |{\csc(\sqrt{\tilde{E}} \, {\pi}/{2})}|$  (positive derivative parts only) corresponding to the odd parity solutions. 
 } 
\label{fig:quantumwellsqrt}
\end{figure}

In summary, we have the following result: the allowed couples (potential depth, energy levels) are given by \eqref{eq:v0even} and \eqref{eq:v0odd} 
or, equivalently, by 
\begin{equation}\label{eq:v0e0sqrt}
\sqrt{\tilde{V}_0}={\sqrt{\tilde{E}}}\, \left|{\sec(\sqrt{\tilde{E}} \, {\pi}/{2})}\right|, \qquad \sqrt{\tilde{V}_0}={\sqrt{\tilde{E}}}\, \left|{\csc(\sqrt{\tilde{E}} \, {\pi}/{2})}\right|
\end{equation}
for the even and odd parity wave functions, respectively. In the previous equations only the positive derivative parts of the functions have to be considered. The chart is reported in Fig.~\ref{fig:quantumwellsqrt}. 

%
This chart is universal as it can be used for a graphical analysis of the allowed energy levels for arbitrary potential depth and well width. 

\subsection{Example of use of the chart}

Let us consider an electron, for which 
$ {8 m}/{h^2}=2.66\cdot 10^{18} \, [eV^{-1} \, m^{-1}]$, 
in a well of width $w=10 \AA$. The normalizing constant is then ${8 m} w^2 /{h^2}=2.66 \, [eV^{-1}]$. Assume a potential depth $V_0= 4.9 \, [eV]$, 
 giving the normalized potential $\tilde{V}_0=13$. 

From the chart we see that for the ordinate $\sqrt{\tilde{V}_0}=3.6$ there are four possible normalized energy levels, approximately of values $\sqrt{\tilde{E}} \in\{ 0.85,  1.7,  2.5,  3.3\}$. Squaring we have $\tilde{E} \in \{0.72,  2.89,  6.25,  10.89\}$, which in electronvolt are \, $E[eV] \in\{ 0.27 , \, 1.09 , \, 2.36, \, 4.1\}$. 

These values, obtained from the graph, are quite close to the exact values, which can calculated as the numerical solution of \eqref{eq:v0e0sqrt} to be $\tilde{E} \in \{0.72,  2.85,  6.30,  10.73\}$, $E[eV] \in \{0.27 , \, 1.07 , \, 2.37, \, 4.04\}$. 

By inspection of the chart we can also see the effect of an increase or decrease in the potential depth level. 




%

\subsection{Some insights from the chart}

The graph of the allowed couple (potential, energy) reported in Fig.~\ref{fig:quantumwellsqrt} leads simply to several observations. 
Some are reported below, where $k\in \mathbb{N}$ denotes a non-negative integer.
\begin{enumerate}
\item Quantization does not depend separately on the system parameters, but on the product $V_0 m w^2$.
\item For a given normalized potential $\tilde{V}_0$, the overall number of energy levels is $N=\lfloor\sqrt{\tilde{V}_0}\rfloor+1$ \,.
%
%

\item 
There is a solution $E=V_0$ only for specific combinations of $V_0, m, w$, producing an integer $\sqrt{\tilde{V}_0}$. In particular, when $\tilde{V}_0=(2k)^2$ and $\tilde{V}_0=(2k+1)^2$, for even and odd parity, respectively.
%
\item 
As a particular case, if we let $\tilde{V}_0 \to \infty$ we obtain the infinite depth quantum well. 
The vertical asymptotes in the chart are at $\sqrt{\tilde{E}}=k$, corresponding to $\cos(  \sqrt{\tilde{E}} \, {\pi}/{2})=0$ or $\sin(\sqrt{\tilde{E}} \, {\pi}/{2})=0$. 
Thus, the normalized energy levels for the infinite depth well are $\tilde{E}=k^2$, and the unnormalized allowed energy levels are $E=k^2 h^2 / (8 m w^2)$. 
%
%
\end{enumerate}
\section{Conclusions}
For the classic problem of a particle in a finite potential quantum well, the chart in Fig.~\ref{fig:quantumwellsqrt} described in this note gives the possible couples (normalized potential, normalized energy). 
It has the advantages to be universal (i.e., valid for arbitrary system parameters $V_0, m,$ and $w$), and to give general insights on the role of the parameters, on the number of quantized energy levels, and on the relation with the infinite depth well case. The chart is simply obtained by plotting the functions ${\sqrt{\tilde{E}}}\, |{\sec(\sqrt{\tilde{E}} \, {\pi}/{2})}|$ and ${\sqrt{\tilde{E}}}\, |{\csc(\sqrt{\tilde{E}} \, {\pi}/{2})}|$.

\bibliographystyle{IEEEtran}
\bibliography{quantum}

\end{document}